\documentclass{article}
\usepackage[T1]{fontenc} 
\usepackage[utf8]{inputenc} 
\usepackage{ismir,amsmath,cite,url}
\usepackage{graphicx}
\usepackage{color}
\usepackage{array}
\usepackage{amssymb}
\usepackage{makecell} 
\usepackage{xcolor}

\usepackage{lineno}
\usepackage{adjustbox}

\newcommand{\GTSD}{\texttt{HX-D}}
\newcommand{\GTSIH}{\texttt{HX-IH}}
\newcommand{\stoller}{\texttt{ST}}
\newcommand{\vaglio}{\texttt{VA}}
\newcommand{\gupta}{\texttt{GU}}
\newcommand{\emir}{\texttt{EM}}
\newcommand{\na}{\textcolor{gray}{N/A}}

\newcommand{\PreserveBackslash}[1]{\let\temp=\\#1\let\\=\temp}
\newcolumntype{C}[1]{>{\PreserveBackslash\centering}p{#1}}
\newcolumntype{R}[1]{>{\PreserveBackslash\raggedleft}p{#1}}
\newcolumntype{L}[1]{>{\PreserveBackslash\raggedright}p{#1}}

\title{HCLAS-X: Hierarchical and Cascaded Lyrics Alignment System Using Multimodal Cross-Correlation}

\threeauthors
  {Minsung Kang} {
  Gaudio Lab, Inc.\\ {\tt m@gaudiolab.com}}
  {Soochul Park} {Gaudio Lab, Inc.\\ {\tt st@gaudiolab.com}}
  {Keunwoo Choi} {Gaudio Lab, Inc. \\ {\tt k@gaudiolab.com}}

\begin{document}

\maketitle
\begin{abstract}

In this work, we address the challenge of lyrics alignment, which involves aligning the lyrics and vocal components of songs. This problem requires the alignment of two distinct modalities, namely text and audio. To overcome this challenge, we propose a model that is trained in a supervised manner, utilizing the cross-correlation matrix of latent representations between vocals and lyrics.
Our system is designed in a hierarchical and cascaded manner. It predicts synced time first on a sentence-level and subsequently on a word-level. This design enables the system to process long sequences, as the cross-correlation uses quadratic memory with respect to sequence length.
In our experiments, we demonstrate that our proposed system achieves a significant improvement in mean average error, showcasing its robustness in comparison to the previous state-of-the-art model. Additionally, we conduct a qualitative analysis of the system after successfully deploying it in several music streaming services.

\end{abstract}

\section{Introduction}\label{sec:introduction}

\begin{figure*}[ht]
 \centerline{
 \includegraphics[width=2.2\columnwidth]{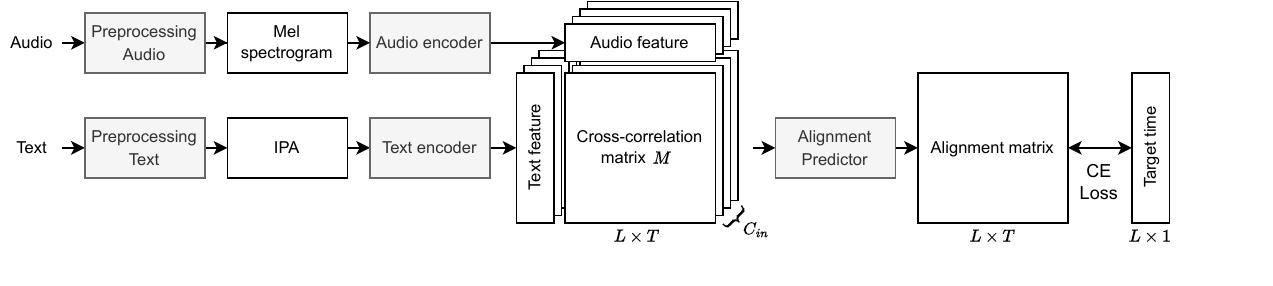}}
 \caption{The block diagram of the proposed system during training process. The CBHG module encodes the preprocessed text and audio into the text features \( \in \mathbb{R}^{C_{in} \times C_{encoder} \times L} \) and the audio features \( \in \mathbb{R}^{C_{in} \times C_{encoder} \times T} \), to create a cross-correlation matrix \( M \in \mathbb{R}^{C_{in} \times L \times T}   \) through matrix multiplication between the two modalities.
 The UNet-based alignment predictor uses this matrix as an input to estimate an alignment matrix. Finally, a CE loss is computed between the estimate and the ground truth alignment vector.
 }
 \label{fig:train}
\end{figure*}

\begin{figure*}[ht]
 \centerline{
 \includegraphics[width=2.1\columnwidth]{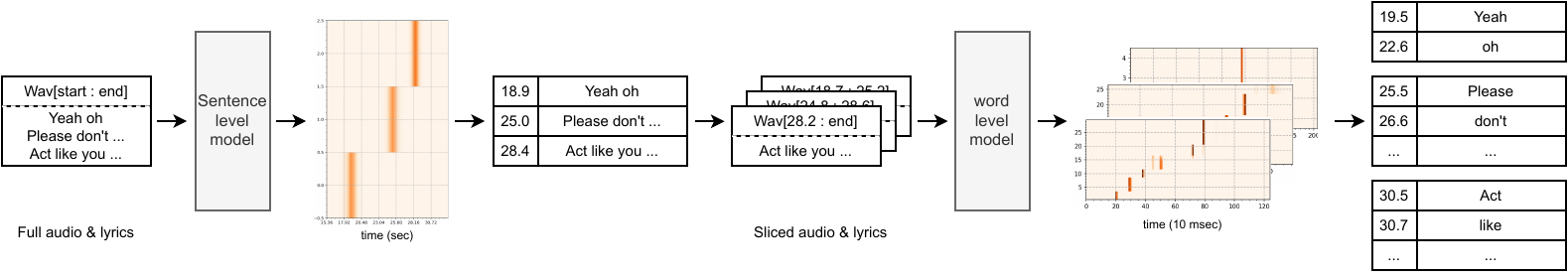}}
 \caption{The block diagram of the proposed system during inference. During the inference process, the sentence-level model is applied to segment the audio into sentence units. Then, the word-level model predicts the timing of each word based on the sliced audio and corresponding lyrics.}
 \label{fig:inference}
\end{figure*}

Lyrics alignment is a task to align music audio (vocals) and the corresponding lyric texts. 
The demands for automating such systems are increasing as music services provide lyrics, while manual alignment remains to be costly. Some music streaming services have a karaoke mode, which necessitates even more precise lyric alignment in a large scale. Lyrics alignment method can be also adopted to synchronize videos and subtitles, whose manual alignment is a tedious job as well. 



Lyrics alignment task is a special case of sequence alignment problems, along with similar popular problems in speech processing and bioinformatics. There have been various methods and studies proposed in many areas, some of which have been adopted to lyrics alignment.  
However, there are some aspects that make lyrics alignment special and difficult.
First, compared to other types of sequences such as speech, singing includes a wide range of distortions that make the alignment a fuzzy problem. For example, there are unclear / modified / omitted pronunciation as well as repeated words and phrases, which are rarely represented in common text. In fact, the repetition turned out to be the most difficult factor in our analysis.
Second, music audio signals are noisy in various senses. Accompaniments reduces the signal-to-noise ratio for the audio part,  singing varies in pitch and duration even for the same word, and the voice can be often heavily modified by audio effects.
Third, in terms of model architecture, the capability to understand temporal context through the entire sequence is important. However, songs are relatively lengthy sequences, making the problem a particularly challenging one from a memory perspective.  

In practice, another issue of developing a high-quality lyrics alignment system is the absence of large and well-organized datasets, which is crucial in the current era dominated by data-driven approaches. 
DALI dataset is well-organized but contains 5358 songs, which is prohibitively small to train modern deep learning models \cite{meseguer2019dali}. DAMP dataset has 6903 songs, but it consists of amateur singers' a cappella, not representing the real popular music \cite{kruspe2016bootstrapping}. The Jamendo evaluation dataset consists of only 20 songs and lacks of recent music.

To address the aforementioned issues, we propose a novel approach. 
First, a cross-correlation matrix is calculated using two latent features and passed to an alignment predictor module to finally predict the alignment time, as illustrated in Figure \ref{fig:train}. Second, we address memory issues by implementing a hierarchical and cascaded alignment pipeline that consists of two stages: sentence-level and word-level -- as shown in Figure \ref{fig:inference}. At the sentence-level stage, we take into account the entire song, which enables us to avoid a local perspective and increase the overall robustness of the alignment process. Subsequently, at the word-level stage, we focus on precisely estimating the timing of each word within a given short segment of the song. Third,  due to the fuzziness of the task discussed earlier, we avoid defining the task as a monotonic alignment and do not rely on any forced alignment algorithms. Finally, we focus on the precise pronunciation of words using the International Phonetic Alphabet (IPA)\cite{ladefoged1990revised, international1999handbook}. This helps reduce the workload of the text encoder and enables the use of a lightweight text encoder module. It also makes the proposed system easily adaptable to various languages.

In Section \ref{sec:related_works}, we introduce the previous approaches for the problem. In Section \ref{sec:methodology}, we describe the details and roles of each module depicted in Figure \ref{fig:train}. In Section \ref{sec:experiments}, we report the experimental setup and results. In Section \ref{sec:discussion}, we discuss the performance, properties, and statistical results of our model. Finally, we conclude our work in Section \ref{sec:conclusion}.

\section{Related works}\label{sec:related_works}

In this section, we will briefly summarize the existing methods in three parts: DTW families, Viterbi families, and the other types. 

Dynamic Time Warping (DTW) is one of forced alignment algorithms \cite{muller2007dynamic}. Intrinsically, to measure the distance between two modalities, the input audio and text must be transformed into the same latent feature space. 
Lee and Scoot \cite{lee2017word} suggest using DTW to align features extracted from synthesized vocal audio from lyrics and original music. 
Chang et al. \cite{chang2017lyrics} employed Canonical Time Warping (CTW) \cite{zhou2009canonical}, which is a more general concept of DTW, for aligning audio and text feature. They utilized Non-negative Matrix Factorization (NMF) to extract audio feature and utilized vowel class tokens as the text feature \cite{lee1999learning}. 

Viterbi algorithm, which is also a type of forced alignment algorithms, has become more popular in recent years \cite{forney1973viterbi}. This is because unlike DTW, it is suitable with ASR models. If you have reference lyrics and a well-trained ASR model that outputs phoneme or character token logits by frames, it can produce the optimal alignment. Some early works in this field such as \cite{fujihara2011lyricsynchronizer, mauch2011integrating} employed the Viterbi algorithm with an HMM-based ASR model\cite{rabiner1989tutorial} to recognize speech from preprocessed audio using various signal processing techniques. In \cite{stoller2019end}, an end-to-end solution with good performance was proposed by using the Wave-U-Net model\cite{stoller2018wave} trained by CTC loss\cite{graves2006connectionist} as an ASR model. 

Based on Stoller's approach, Vaglio et al. employed International Phonetic Alphabet (IPA) for multi-lingual alignment \cite{vaglio2020multilingual} and Gupta et al. utilized genre information as an additional feature to improve the alignment performance \cite{gupta2020automatic}. Demir et al. proposed a novel methodology, which involves using the concept of an anchor and dividing audio into smaller segments for efficient forced alignment \cite{demirel2021low}.

Plenty of the previous studies are based on forced alignment algorithms such as Viterbi algorithm \cite{forney1973viterbi} because of the similarity between lyrics alignment and other sequence alignment tasks~\cite{chang2017lyrics, stoller2019end, muller2007dynamic, zhou2009canonical}.
However, there is a critical limitation to use forced alignment algorithms for our task. 
In those approaches, an error that once occurs in a possibly difficult part may cause more failures in the surrounding time steps, leaving a critical negative impact on the overall performance.
To overcome the limitation, in this work, we propose an end-to-end supervised training model that shows strong prediction performance, fast inference speed, and multi-language compatibility.

\section{Methodology}\label{sec:methodology}

\subsection{Preprocessing Text}\label{subsec:preprocess_text}
\begin{figure}[ht]
 \centerline{
 \includegraphics[width=0.8\columnwidth]{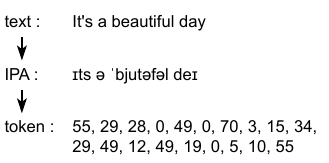}}
 \caption{Example of preprocessing text steps}
 \label{fig:ipa}
\end{figure}

The lyrics are provided in a textual format, segmented by a new line character.
We convert this text into IPA tokens. An example of this conversion is shown in Figure \ref{fig:ipa}. This conversion enables the text encoder module to focus on pronunciation. 
Since any language can be represented with IPA tokens, this design choice enables the system to be compatible with any languages even if they are not included in the training data.

We utilize two Python libraries, \texttt{eng-2-ipa},\footnote{\url{https://github.com/mphilli/English-to-IPA}} and \texttt{kopron}\footnote{\url{https://github.com/kord123/ko_pron}}, to convert English and Korean lyrics in all our training data. As a result, we obtain a vocabulary size of 76 tokens -- 75~IPA symbols and a special token for silence. When processing languages that are not English or Korean, there may be some IPA tokens that are not part of the chosen 75~IPA symbols.
In that case, we replace them with the most similar available pronunciation using a table in \cite{ladefoged1990revised} that proposes the similarity between two pronunciations.

\subsection{Preprocessing Audio}\label{subsec:preprocess_audio}
We first separate the vocal from the music so that the model can focus on the relevant information to the task. We utilize our own voice separation model,\footnote{\url{https://studio.gaudiolab.io/}}
which achieves a 10.0 SDR on the MUSDB18 dataset. The audio of the separated vocal is then transformed into a mel spectrogram and used as an audio input feature to the audio encoder. The audio is sampled at a rate of 16~kHz. Mel spectrogram is configured with a window size of 1024, a hop size of 512, and a mel filter bank of 80.

For the sentence-level alignment, we add another step to shorten the sequence length by 4~times. We compress 4~consecutive frames into a single frame by concatenating them along the frequency axis.

\subsection{Text and Audio Encoder Modules}\label{subsec:model_architecture}
We choose a lightweight encoder, instead of generally more powerful architectures such as Transformers \cite{vaswani2017attention}, because the encoder does not need a long-range context information. Our choice for both text and audio encoders is CBHG module used in \cite{wang2017tacotron, lee2017fully}. The CBHG module is an elaborately designed architecture that combines 1D Convolutional Banks, Highway Networks, and Gated Recurrent Units to extract representations from long sequences. This is achieved by leveraging the strengths of each component: the 1D convolutional layers observe consecutive sequences, the Highway network selectively amplifies important features, and the final GRU layer captures a sequential representation.

At the end of CBHG, a linear layer is added to increase the number of channels by a factor of \( C_{in}\), followed by a reshape operation. The resulting encoder output shape is represented by a tensor with dimensions \( C_{in} \times C_{encoder} \times (L \ \text{ or } \ T)\). \(C_{encoder}\) denotes the number of channels of CBHG, \(L\) represents the length of sentence or the length of IPA, and \(T\) means the number of time frames.

By matrix multiplication applied to the resulting encoded latent representations along the \( C_{encoder} \) axis, we can get cross-correlation matrix \(M \in \mathbb{R} ^{ C_{in} \times L \times T} \). This is then used as input of the alignment predictor module.

\subsection{Alignment Predictor Module}
We employ a UNet model with GRU. The initial 2-dimensional convolution layer outputs a 32-channel tensor. The channel size is subsequently doubled after each pooling layer. At each depth level, there are two convolutional layers, and skip connections between the down- and up-sampling structures.

The alignment predictor module takes the previously calculated \(M\) as an input to output a tensor -- alignment matrix -- with a shape of \(L \times T\). It then applies the argmax function along the time axis to make a time prediction for each text. 

The cross-correlation matrix before the alignment predictor results in suboptimal alignment outcomes, particularly when the song contains repetitive phrases. If the encoders extract a proper pronunciation latent from each input, the alignment predictor module searches for the most suitable alignment from the cross-correlation matrix.

\subsection{Hierarchical and Cascaded Architecture}
As illustrated in Fig \ref{fig:inference},  our approach involves training two separate models: a sentence-level model and a word-level model. Although these models share a similar architecture, their hyperparameter settings differ slightly, as outlined in Table \ref{tab:hyperparameter}.

The sentence-level model consists of two 256-dimensional CBHG layers and a UNet with a depth of~4, containing a total of 42.7 million parameters. The word-level model, with 46.9~million parameters, features two 512-dimensional CBHG layers and a UNet with a depth of~3.

\begin{table}[ht]
\begin{tabular}{c|c|c}
                                                         & \multicolumn{1}{c|}{Sentence-level} & \multicolumn{1}{c}{Word-level} \\ \hline \hline
Encoder                                                  & 2 CBHG, 256-dim                     & 2 CBHG, 512-dim                \\ \hline
\begin{tabular}[c]{@{}c@{}}UNet \\ Channels\end{tabular} & 32, 64, 128, 256              & 32, 64, 128             
\end{tabular}
\caption{A detailed description of the model parameter settings.}
\label{tab:hyperparameter}
\end{table}

\subsection{Loss Function}
We utilize the cross-entropy (CE) loss function for training along the time axis. We also tested L1 and L2 loss in the preliminary experiment. Although L1 loss is more directly correlated with the evaluation metrics than other losses, the CE loss yielded the best results.

\begin{table*}[ht]
 \centering
\begin{tabular}{C{2.5cm} | C{2cm} | C{2cm} | C{2cm} | C{2cm} | C{2cm}}

 &
  \textbf{MAE $\downarrow$} &
  \textbf{MedAE $\downarrow$} &
  \textbf{Perc $\uparrow$} &
  \textbf{\(\text{Mauch}_{0.3} \) $\uparrow$} &
  \textbf{\(\text{Mauch}_{0.2} \) $\uparrow$} \\ \hline
  \hline
{\stoller{}}\cite{stoller2019end}   & 0.38  & 0.097 & 76.8\% & 0.87 & 0.82\\ \hline
{\vaglio{}}\cite{vaglio2020multilingual}   & 0.37  & \na  & \na & 0.92 & \na \\ \hline
{\gupta{}}\cite{gupta2020automatic}   & 0.22  & 0.050 & \na & \textbf{0.94} & \na \\ \hline
{\emir{}}\cite{demirel2021low}   & 0.31  & 0.050 & \na & 0.93 & \na \\ \Xhline{1.2pt}  
{\GTSD{}} & 0.42  & \textbf{0.043}  & 83.7\% & 0.89 & 0.87 \\ \hline
\GTSIH{}   & \textbf{0.16}  & \textbf{0.043} & \textbf{89.0\%} & 0.93 & \textbf{0.91} \\ 
\end{tabular}
 \caption{Evaluation result on Jamendo dataset. \GTSD{} and \GTSIH{} represent our model trained by DALI and in-house dataset respectively. The existing model names, namely \stoller{}, \vaglio{}, \gupta{}, and \emir{} are derived from the first and second letters of the first author's name. The results of all metrics are evaluated for each song and then aggregated by taking the average.}
 \label{tab:Jamendo}
\end{table*}

\section{Experiments}\label{sec:experiments}

\subsection{Dataset}\label{subsec:dataset}
We use two datasets, the DALI dataset and an in-house dataset, to obtain two different trained systems. 
The DALI dataset consists of 5358 songs, of which 4590 were used as training data. The dataset includes onset and offset at the levels of notes, words, lines and paragraphs. Only onset at the line and word levels is used for training. The in-house dataset consists of approximately 67,000 Korean and English language songs, each with a corresponding pair of song and lyrics. The dataset also includes start times for sentence-level markings. 

For evaluation purpose, we use the Jamendo dataset from MIREX 2019 and the Mandarin pop song dataset from MIREX 2018. The Jamendo dataset has start time of word-level, but Mandarin pop song dataset only includes sentence-level start time, making the evaluation limited to the sentence-level module. We still use the former dataset to assess the generalizability towards music with a different language and genres.


\subsection{Evaluation Metric}\label{subsec:body}
We follow \cite{stoller2019end} and use these four metrics -- Mean Absolute Error (MAE), Median Absolute Error (MedAE), Perc, and Mauch metric\cite{mauch2011integrating}. Perc refers to the percentage of correct duration in the total duration. Mauch metric measures the ratio of correctly predicted start times to the true bounds, given a value of \(\tau\).
These are all song-level metrics and we compute the average to aggregate them within a dataset. 

\begin{equation}\label{eq:MAE}
\text{MAE} = \dfrac{\sum_{w=1}^{W} \mid t_{pred}^w - t_{ref}^w \mid } {W}
\end{equation}

\begin{equation}\label{eq:MedAE}
\text{MedAE} = \underset{1 \leq w \leq W}{\mathrm{median}}\big( \mid t_{pred}^w - t_{ref}^w \mid \big)
\end{equation}

\newcommand\scalemath[2]{\scalebox{#1}{\mbox{\ensuremath{\displaystyle #2}}}}
\begin{equation}\label{eq:Perc}
\text{Perc} = \scalemath{0.77}{\dfrac{\sum_{w=1}^{W} \max\Big( \min{ \big(e_{ref}^w, e_{pred}^w} \big) - \max{ \big( t_{ref}^w, t_{pred}^w \big) }    , 0 \Big)  } {Duration}}
\end{equation}

\begin{equation}\label{eq:Mauch}
\text{Mauch}_{\tau} = \dfrac{\sum_{w=1}^{W}\mathbf{1}_{\mid t_{pred}^w - t_{ref}^w  \mid < \tau }}{W}
\end{equation}

In the equations, \(W\) refers to the total number of words in a song and \(w\) refers to word index. 
The terms \(t_{ref}^w , t_{pred}^w\) correspond to the reference and predicted start time of the \(w\)-th word, respectively. The value \(e^w\) represents the end time of a word. 
However, in the absence of this information in the Jamendo dataset, we set \(e^w \) equals to \(t^{w+1} \).

\subsection{Experiments Details}\label{subsec:exp_details}
 The sentence-level model was trained with a batch size of 24, a learning rate of 5e-4, and a weight decay of 1e-3. To improve model performance, we build an ensemble of four and eight models for \GTSD{} and \GTSIH{} respectively, and use the average of the logits. We use data augmentation using Audiomentation library when training of the DALI-only model.\footnote{https://github.com/iver56/audiomentations} In total, five augmentation components are employed, namely AddGaussianNoise, PitchShift, PolarityInversion, BandPassFilter, and Gain.

 The word-level model is trained using a batch size of 64, a learning rate of~5e-4, and a weight decay of~1e-7. When we train the word-level model in \GTSIH{}, we utilize Noisy Student Training (NST) \cite{xie2020self} with three iterations on the in-house dataset.

\subsection{Evaluation on Jamendo dataset}\label{subsec:eval_jamendo}
Before examining the results of the Jamendo dataset, we would like to inform you about the bias that is added to the predicted time during evaluation. The \stoller{} method uses a bias of 0.18 seconds, the \GTSD{} uses a bias of 0.05 seconds, and the \GTSIH{} uses a bias of 0.1 seconds for evaluating the Jamendo dataset.


Table \ref{tab:Jamendo} shows that \GTSIH{} model outperforms all other models across all metrics except for the Mauch metric. While the \GTSD{} model shows the lowest MAE, it outperforms all the previous systems in terms of MedAE. This indicates that both of our models have similar performance overall (as their median scores being similar), but \GTSD{} model would have some large errors on relatively more difficult cases. 

\begin{figure}[]
 \centerline{
 \includegraphics[width=1\columnwidth]{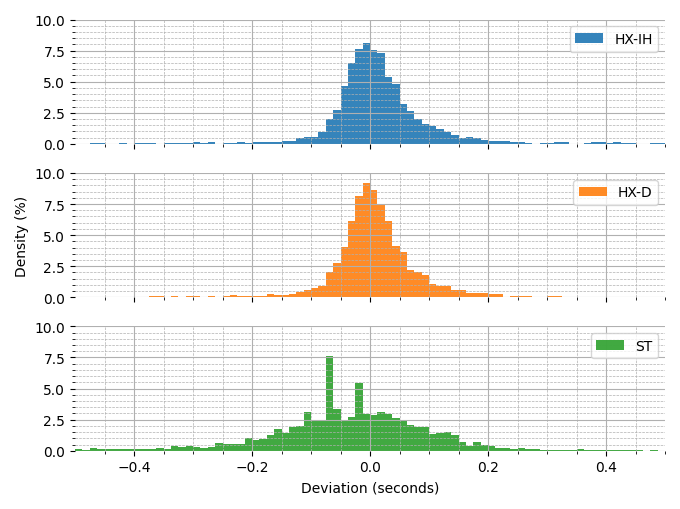}}
 \caption{Deviation histograms of three models on the Jamendo dataset. (\(\text{Deviation} = t_{pred} - t_{ref}\))  }
 \label{fig:histogram_deviation}
\end{figure}

The Mauch metric represents the true/false ratio with a tolerance range of $\tau$. It is related to the confusion matrix, which we will discuss in Section \ref{sec:discussion}. In Table \ref{tab:Jamendo}, $\text{Mauch}_{0.3}$ scores are quite similar among the \GTSIH{}, \GTSD{} and \stoller{}. However, the \stoller{} has a $\text{Mauch}_{0.2}$ value that decrease steeply to 0.82, while the \GTSD{} has the value of 0.87 and the \GTSIH{} has the value of 0.91. Figure \ref{fig:histogram_deviation} provides an explanation for this result, showing the histogram of deviation for each model. Our two models have almost identical histogram shapes, which are more centered, in contrast to the histogram of \stoller{}. 
This indicates that our models are more suitable if one sets tight tolerance range to ensure high-quality alignment.

\begin{table}[]
 \begin{center}
\begin{tabular}{c|c|c|c}
\hline
 &
  \textbf{MAE $\downarrow$} &
  \textbf{MedAE $\downarrow$} &
  \textbf{Perc $\uparrow$} \\ \hline
\GTSD & 0.10 & 0.09 & 98.2\% \\ \hline
\GTSIH & 0.12 & 0.09 & 98.2\% \\ \hline
\end{tabular}
\end{center}
 \caption{An evaluation on the Mandarin pop song dataset, consisting of 20 songs that contain only vocal tracks and Chinese character lyrics. Pinyin text files are not used in this evaluation.}
 \label{tab:Mandarin}
\end{table}

\subsection{Evaluation on Chinese Pop Song dataset}\label{subsec:eval_mandarin}
Although our dataset primarily comprises Korean and English songs, we achieved favorable results on the Mandarin pop song dataset in MIREX 2018. 
We believe that this language generalizability is thanks to the adoption of IPA in our model. 
As shown in Table \ref{tab:Mandarin}, we achieved strong results with an MAE of 0.12 seconds and a MedAE of 0.09 seconds, which is not too different from English song results in Table \ref{tab:Jamendo}. Although we could not perform objective word-level evaluation, we found our model demonstrates satisfactory word-level alignment performance when tested internally on several other languages including Chinese, Japanese, French, and Spanish.

\begin{table}[]
\centering
\begin{tabular}{C{3.4cm}|C{1.6cm}}
Text preprocessing   & 2.64 sec \\ \hline
Audio preprocessing  & 1.33 sec \\ \hline
Sentence-level model & 0.74 sec \\ \hline
Word-level model     & 0.11 sec \\ \hline \hline
total                & 4.82 sec \\
\end{tabular}%
\caption{Average processing time of \GTSIH{}. We tested with the 20~songs from the Jamendo dataset on one NVIDIA RTX 3090 GPU}
\label{tab:processing_time}
\end{table}

\subsection{Processing time}\label{subsec:processing_time}
Table \ref{tab:processing_time} shows our alignment system can perform the task at an average speed of 4.8 seconds per song without considering offset time such as model loading. In our analysis, text preprocessing occupies nearly half of the total processing time. Audio preprocessing takes an average of 1.33 seconds per song which, as discussed in section \ref{subsec:preprocess_audio}, includes vocal separation and processing of mel~spectrograms from waveforms. Our proposed sentence-level module and word-level module are significantly lighter in processing time compared to the aforementioned preprocessing modules.

\begin{figure}[]
 \centerline{
 \includegraphics[width=1\columnwidth]{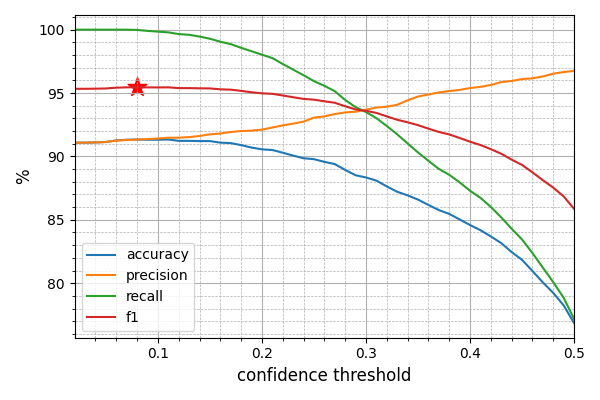}}
 \caption{Various statistical terms of proposed model trained by in-house dataset when the absolute error true bound is 0.2 seconds.}
 \label{fig:Jamendo_analysis}
\end{figure}

\begin{table}[]
\centering
\resizebox{\columnwidth}{!}{%
\begin{tabular}{c|c|ccc|c|c}
h=0.08 & \textbf{True} & \textbf{False} &   & h=0.32 & \textbf{True} & \textbf{False} \\ \cline{1-3} \cline{5-7} 
\textbf{Accept} & 91.07\%      & 8.65\%        &   & \textbf{Accept} & 84.16\%      & 5.44\%        \\ \cline{1-3} \cline{5-7} 
\textbf{Reject} & 0.02\%       & 0.26\%        &   & \textbf{Reject} & 6.92\%       & 3.47\%        \\ 
\end{tabular}%
}
\caption{Confusion matrix of proposed model trained by in-house dataset when the absolute error true bound is 0.2 seconds and confidence thresholds are 0.08 and 0.32.}
\label{tab:TFtable}
\end{table}

\section{Discussion}\label{sec:discussion}
As explained in Section~2, there are various difficulties in solving the lyrics alignment problem. The issues include repeated phrases, inaccuracies in audio-text pairs, and inaccuracies in pronunciation due to various musical reasons. Even in the Jamendo dataset, of which the items are much easier to sync lyrics than actual popular music, we can only achieve approximately 91\% accuracy with a tolerance range of $\pm$ 0.2 seconds. While this result is a better result than the previous models, it is still unsatisfactorily low for real-world use cases such as streaming or karaoke applications. Hence, we apply additional inspection process for quality assurance.

During the inspection process, the concept of confidence for predicted alignment becomes crucial, as it informs potential candidates for correction. Deterministic alignment algorithms, on the other hand, are of no help for the QA process. In addition, visualizing the prediction, as shown in Figure \ref{fig:inference}, can significantly reduce the time required to identify potential errors.

The maximum F1 score achieved is 95.17\%, with a recall of 98.57\% when the confidence threshold is set to 0.08, as shown in Figure \ref{fig:Jamendo_analysis}. The resulting confusion matrix is presented in Table \ref{tab:TFtable}. Alternatively, one may choose to manually verify a certain portion of the alignments. For example, to manually verify the items with bottom 10\% confidence, the confidence threshold would be set to 0.32. We can then evaluate the remaining 90\% of items as that means the performance of the automatically aligned items. We would obtain a false positive of 5.44\% and F1 score of 93.16\%, resulting in a 17\% reduction in terms of MAE, as seen in Table \ref{tab:TFtable}.

Based on these benefits, our end-to-end alignment system is not only efficient and high-performing but also enabling an efficient manual inspection. This is important because once deploying the system, one is able to i) improve the model by continual active learning with the failure cases or ii) fix the ground truth.

\section{Conclusion}\label{sec:conclusion}
In this paper, we propose a novel end-to-end approach for aligning lyrics with audio. Our system achieves state-of-the-art performance in absolute error with fast inference speed and is capable of processing multiple languages. Compared to forced alignment algorithms, the proposed system is non-deterministic, which enables clients to easily detect potential inaccuracies. It is also more resilient to catastrophic failures. Our model architecture can be adopted to other alignment problems by adapting the encoder modules to the corresponding data types. 

Like any other machine learning systems, our approach is heavily reliant on the quality and diversity of the training data. Therefore, we believe that further research in creating more diverse and realistic datasets for alignment problems is crucial for evaluating and improving the accuracy of such systems. We also hope that our work inspires more progress in this field, leading to the development of more robust and adaptable alignment models that can be applied to various real-world scenarios.


\bibliography{GTS}

%
%
%
%
%

\end{document}